\documentclass[aip,jap,reprint]{revtex4-1} 
\usepackage{graphicx}

\begin{document}

\title{Anomalous Hall effect in the noncollinear antiferromagnet Mn$_5$Si$_3$} 

\author{Christoph S\"urgers}
\email[]{christoph.suergers@kit.edu}
\affiliation{Physikalisches Institut, Karlsruhe Institute of Technology, P.O. Box 6980, 76049 Karlsruhe, Germany}
\author{Wolfram Kittler}
\affiliation{Physikalisches Institut, Karlsruhe Institute of Technology, P.O. Box 6980, 76049 Karlsruhe, Germany}
\author{Thomas Wolf}
\affiliation{Institut f\"ur Festk\"orperphysik, Karlsruhe Institute of Technology, P.O. Box 3640, 76021 Karlsruhe, Germany}
\author{Hilbert v. L\"ohneysen}
\affiliation{Physikalisches Institut, Karlsruhe Institute of Technology, P.O. Box 6980, 76049 Karlsruhe, Germany}
\affiliation{Institut f\"ur Festk\"orperphysik, Karlsruhe Institute of Technology, P.O. Box 3640, 76021 Karlsruhe, Germany}

\date{\today}

\begin{abstract}
Metallic antiferromagnets with noncollinear orientation of magnetic moments provide a playground for investigating spin-dependent transport properties by analysis of the anomalous Hall effect. The intermetallic compound Mn$_5$Si$_3$ is an intinerant antiferromagnet with collinear and noncollinear magnetic structures due to Mn atoms on two inequivalent lattice sites. Here, magnetotransport measurements on polycrstalline thin films and a single crystal are reported. In all samples, an additional contribution to the anomalous Hall effect attributed to the noncollinear arrangment of magnetic moments is observed. Furthermore, an additional magnetic phase between the noncollinear and collinear regimes above a metamagnetic transition is resolved in the single crystal by the anomalous Hall effect. 
\end{abstract}


\maketitle 

\section{Introduction}
Today's spintronic devices mostly utilize the spin-dependent electronic transport in or between ferromagnets. For a long time, antiferromagnetic materials with zero net magnetization have rarely been considered to be of interest for information storage due to the difficulty to externally manipulate or read-out the magnetization state. However, quite recently, antiferromagnetic metal spintronic devices have been proposed to show current-induced phenomena like in ferromagnets such as spin-transfer torque \cite{macdonald_antiferromagnetic_2011,cheng_spin_2014}, spin pumping \cite{cheng_spin_2014}, and domain-wall motion \cite{tveten_staggered_2013}. This opens up new opportunities in creating functional devices based on antiferromagnets as active components \cite{jungwirth_antiferromagnetic_2015}. In addition, zero stray fields, reduced switching currents, and ultrafast switching by light \cite{kimel_inertia-driven_2009} are further advantages specific to antiferromagnets. 
\\
In solid-state electronic transport, the Hall effect takes a unique position because it probes electronic states directly at the Fermi level and the normal part of the Hall effect provides for single-band metals a measure of the Fermi volume, i.e., the volume in momentum space enclosed by the Fermi surface. At the onset of magnetic order additional contributions come into play and the measured Hall resistivity $\rho_{xy} = V_yt/I_x$ ($V_y$: Hall voltage across the sample width, $I_x$: current along direction x, $t$: sample thickness) comprises, in addition to the ordinary term arising from the Lorentz force acting on the charge carriers, a contribution called extraordinary or anomalous Hall effect (AHE)
\begin{equation}
\rho_{xy} = R_0B_z + \rho_{xy}^{AHE}
\label{eq1}
\end{equation}
where $B_z = \mu_0[H_z+(1-N_z)M_z]$, and $H_z$, $M_z$, and $N_z$ are the magnetic field, magnetization, and demagnetization factor, respectively, along the \textit{z} direction perpendicular to the \textit{xy} plane. Equation \ref{eq1} is valid for low magnetic fields, where the cyclotron frequency $\omega_c$ is much smaller than the mean scattering rate $\tau^{-1}$ of charge carriers, corresponding to $R_0B \ll \rho_{xx}$. 
\\
In long-range ordered ferromagnets, the AHE is a consequence of the broken time-reversal symmetry and spin-orbit coupling (SOC) \cite{nagaosa_anomalous_2010}. The latter leads to  different scattering directions for spin-up and spin-down charge carriers which, together with the spin imbalance, creates a charge accumulation at opposite edges of the sample and a transverse electric field, i.e., a Hall voltage. The AHE is thought to arise from an extrinsic part attributed to energy-dissipative scattering and a disipationless intrinsic part arising from the Berry-phase curvature of the Bloch states in momentum space \cite{jungwirth_anomalous_2002,nagaosa_anomalous_2010}. The individual contributions to the measured Hall voltage depend on the amount of scattering. The Berry-phase concept provides a link between the electronic band structure and the magnitude of the AHE \cite{nagaosa_anomalous_2010}. In ferromagnets, SOC gives rise to a fictitious local field corresponding to a Berry potential. The scattering-independent Berry-phase contribution is more important for the AHE than for any other electronic transport coefficient. Recent studies \cite{miyasato_crossover_2007} revealed a crossover from the extrinsic (skew-scattering) region to the intrinsic region with increasing resistivity $\rho_{xx}$. Scaling the AHE provides a connection between the magnetic and transport properties, enabling the universality class of the phase transition to be reliably determined from the AHE with a prospect of studying the role of dimension on the critical behavior \cite{jiang_scaling_2010}.

\subsection{Hall effect in noncollinear magnetic structures}
Early descriptions \cite{turov_1965} of the AHE in an antiferromagnet (AFM) with two collinear sublattices of magnetization \textbf{M$_1$} and \textbf{M$_2$} considered a Hall effect proportional to the antiferromagnetic vector \textbf{L = M$_1$ - M$_2$}. More recently, the AHE has been investigated in various systems with nontrivial arrangements of the magnetic moments. In these instances, the AHE was explained by the accumulated Berry phase of the electron when moving through the spatially varying magnetization along its path, even without SOC. The remaining contribution to the AHE arising from the magnetic structure and not from SOC has been often related to a topological Hall effect (THE). A THE due to the Berry phase in \textit{momentum space} was observed in Pr$_2$Ir$_2$O$_7$ and Nd$_2$Mo$_2$O$_7$ with pyrochlore structure where the localized spins do not exhibit long-range magnetic order \cite{taguchi_spin_2001,machida_time-reversal_2010}. In the highly correlated metal UCu$_5$, a chirality-induced 'geometrical Hall effect' independent of SOC has been shown to occur due to antiferromagnetically-coupled localized 5\textit{f}-electron spins \cite{ueland_controllable_2012}. In the helimagnets MnGe, MnSi, or FeGe hosting a skyrmion lattice, a THE is due to the winding of the spin texture and the Berry potential in \textit{real space} \cite{nagaosa_topological_2013}. In fact, in this case the THE is a hallmark of skyrmion formation and has been used to reveal depinning and motion of skyrmions \cite{schulz_emergent_2012}. Emergence of a large AHE has also been predicted on the surface of an Fe monolayer on Ir(001) \cite{hoffmann_topological_2015}.
\\
In contrast to ferromagnets where it is often assumed that the anomalous part of the AHE depends linearly on $M$, $\rho_{xy}^{AHE} = S_H \rho_{xx}^2 M$, a vanishing AHE is expected for an AFM with zero net magnetization. However, it has been shown that the intrinsic part of the AHE due to the Berry-phase curvature is not zero if certain symmetries are broken, for instance due to an applied magnetic field or SOC \cite{chen_anomalous_2014}. Although in a collinear AFM the total symmetry recovers and no AHE is generated, a nonzero AHE occurs in a noncollinear magnetic arrangement independent of the strength of SOC which can even be zero. The THE due to the strongly noncollinear magnetization texture should vanish when a collinear magnetic state is induced by a magnetic field $H > H_c$. Alternatively, the AHE of an AFM should strongly change when the spin structure changes from collinear to noncollinear in dependence of temperature \cite{chen_anomalous_2014}. 
\\
In this respect, hexagonal Mn$_3$Ge and cubic Mn$_3$Pt or Mn$_3$Ir have been proposed to show a large AHE due to the nontrivial spin structure \cite{kubler_non-collinear_2014,chen_anomalous_2014,gomonay_berry-phase_2015}. Furthermore, antiferromagnetic Mn$_5$Si$_3$ is an interesting metallic compound which does show collinear and noncollinear phases at different temperatures and can thus serve as model system to study the different contributions to the AHE. In fact, an extra contribution to the AHE caused by the noncollinear magnetic structure and attributed to a THE has been reported for Mn$_5$Si$_3$ films \cite{surgers_large_2014}. Noncollinearity may also be stabilized in the isostructural ferromagnet Mn$_5$Ge$_3$ by uniaxial distortion \cite{stroppa_competing_2007}. Mn$_5$Ge$_3$ and Mn$_5$Ge$_3$C$_x$ ($x \approx 1$) have been proposed as ferromagnetic electrodes for spintronic applications due to their ability to grow epitaxially on Si and GaAs substrates  \cite{zeng_epitaxial_2003,surgers_magnetic_2008,slipukhina_simulation_2009,thanh_epitaxial_2013,fischer_hanle-effect_2014}. In all these Mn compounds the different magnetic structures originate from the sensitivity of the Mn moment on the local atomic environment, in particular the Mn-Mn distances which are found to be the major factor leading to different site-dependent Mn moments \cite{forsyth_spatial_1990}.  
\\
Here, we report on measurements of the electrical resistivity and AHE of a Mn$_5$Si$_3$ single crystal and polycystalline thin films. The latter show indications of a THE in the noncollinear phase which is absent in the collinear phase and for large magnetic fields. A similar behavior is found for the single crystal albeit with a hysteresis in the AHE. Moreover, the metamagnetic transition at high fields gives rise to a strong contribution to the AHE which is absent in the polycrystalline films. Comparison of the data allows to reveal the contributions from different magnetic phases to the Hall effect, such as a metamagnetic transition, and suggests an intermediate magnetic phase between the noncollinear and collinear phase in this compound. 

\subsection{Antiferromagnetic Mn$_5$Si$_3$}
\begin{figure}
\includegraphics[width=0.7\columnwidth]{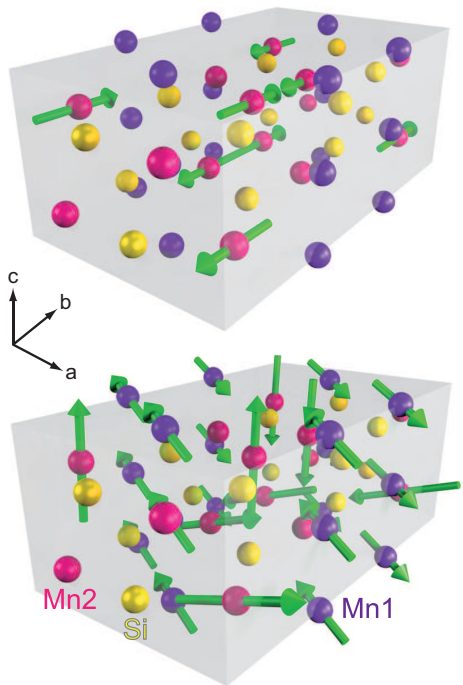}
\caption{\label{fig1}Magnetic structure of Mn$_5$Si$_3$ (orthorhombic unit-cell) in the collinear AF2 phase (top) and in the low-temperature noncollinear AF1 phase (bottom) obtained from neutron scattering  \cite{brown_low-temperature_1992}.}
\end{figure}

In the paramagnetic state, the hexagonal unit cell (space group P6$_3$/mcm) with lattice parameters $a_{\rm h}$ and $c_{\rm h}$ contains Mn$_1$ and Mn$_2$ atoms on two inequivalent lattice sites. An effective magnetic moment of 3.6 $\mu_{\rm B}$ was determined from the paramagnetic susceptibility \cite{gottschilch_study_2012}. The occurrence of long-range antiferromagnetic order below the N\'{e}el temperature $T_{\rm N2}$ is accompanied by a distortion of the orthorhombic unit cell with lattice parameters $a_{\rm r} \approx a_{\rm h}$, $c_{\rm r} \approx c_{\rm h}$, and $b_{\rm r} < \sqrt{3} a_{\rm h}$ \cite{brown_low-temperature_1992,brown_antiferromagnetism_1995,gottschilch_study_2012}. In the AF2 phase between $T_{\rm N2}$ and $T_{\rm N1}$, the Mn$_1$ and one third of the Mn$_2$ atoms do not exhibit an ordered magnetic moment. The remaining Mn$_2$ atoms have magnetic moments $\mu \approx 1.5\, \mu_{\rm B}$ oriented parallel and antiparallel to the crystallographic \textit{b} axis of the orthorhombic cell in a collinear fashion \cite{brown_antiferromagnetism_1995,gottschilch_study_2012}, see Fig. \ref{fig1}.
\\
In the low-temperature antiferromagnetic AF1 phase below $T_{\rm N1}$, the magnetic structure has monoclinic symmetry with atomic positions that can still be described with orthorhombic symmetry of the Ccmm space group without inversion symmetry \cite{brown_low-temperature_1992,silva_magnetic_2002,gottschilch_study_2012}. In this phase, the Mn$_1$ atoms acquire a moment presumably due to the expansion of the lattice along the crystallographic \textit{c} axis and the accompanied increase of the Mn$_1$-Mn$_1$ distance  \cite{brown_low-temperature_1992,silva_magnetic_2002,gottschilch_study_2012,brown_antiferromagnetism_1995}. The moments point into different directions forming a highly non-collinear antiferromagnetic structure, see Fig. \ref{fig1}. One third of the Mn$_2$ atoms still do not exhibit an ordered magnetic moment below $T_{\rm N1}$. The non-collinearity is attributed to frustration \cite{brown_low-temperature_1992,brown_antiferromagnetism_1995}. $T_{\rm N2}$\, = 100 K and $T_{\rm N1}$\,= 62 K have been also determined from susceptibility measurements, and indications of a third anomaly with hysteresis between 30 and 60 K have been found \cite{gottschilch_study_2012}. In that work, a non-collinear phase has been suggested for $T < T_{\rm N1}$\, as well, however with a modified spin arrangement \cite{gottschilch_study_2012}.  

\section{Experimental}
A 160-nm thick Mn$_5$Si$_3$ film was prepared by magnetron sputtering from elemental targets on a sapphire substrate heated to 470 $^{\circ}$C. A mechanical mask was used to obtain a Hall-bar configuration \cite{gopalakrishnan_electronic_2008}. In addition, we discuss data obtained on a 40-nm film which have been published earlier \cite{surgers_large_2014}. The hexagonal structure of the polycrystalline film was confirmed by x-ray diffraction. Sputtered films prepared under the same conditions have a coarse-grained morphology with a grain size less than 100 nm \cite{gopalakrishnan_electronic_2008}. 
\\
The Mn$_5$Si$_3$ crystal was obtained by a combined Bridgman and flux-growth technique using a Mn-rich self flux and a low cooling rate of 1.2$^{\circ}$C/h and was characterized by powder x-ray diffraction as well, again confirming the formation of the Mn$_5$Si$_3$ phase. A thin cuboid piece ($0.9 \times 3.9$\ mm$^2$) of 0.47 mm thickness with the crystallographic $a_{\rm h}$ and $c_{\rm h}$ axes aligned parallel to the long and short edge, respectively, and perpendicularly to the sample normal ($z$ direction) was obtained after orientation by Laue diffraction. Resistivity and Hall-effect measurements were performed in a physical-property measurement system (PPMS) with the field oriented along the $z$ direction perpendicular to the sample $xy$ plane and with the current in plane. Hence, in the orthorhombic phase the magnetic field is along the $b_{\rm_r}$ direction. Data were taken for both field directions and were symmetrized by $\rho_{xy}(H) = [\rho_{xy}(+H)-\rho_{xy}(-H)]/2$ for each field and by taking into account the direction of the measurement loop. Magnetization curves of the film and of the single crystal were acquired in a vibrating sample magnetometer up to 12 T and in a SQUID magnetometer up to 5 T with the field applied in the same orientation as for the Hall-effect measurements.

\section{Results}  
\subsection{Thin films} 

\begin{figure}
\includegraphics[width=0.8\columnwidth,clip=]{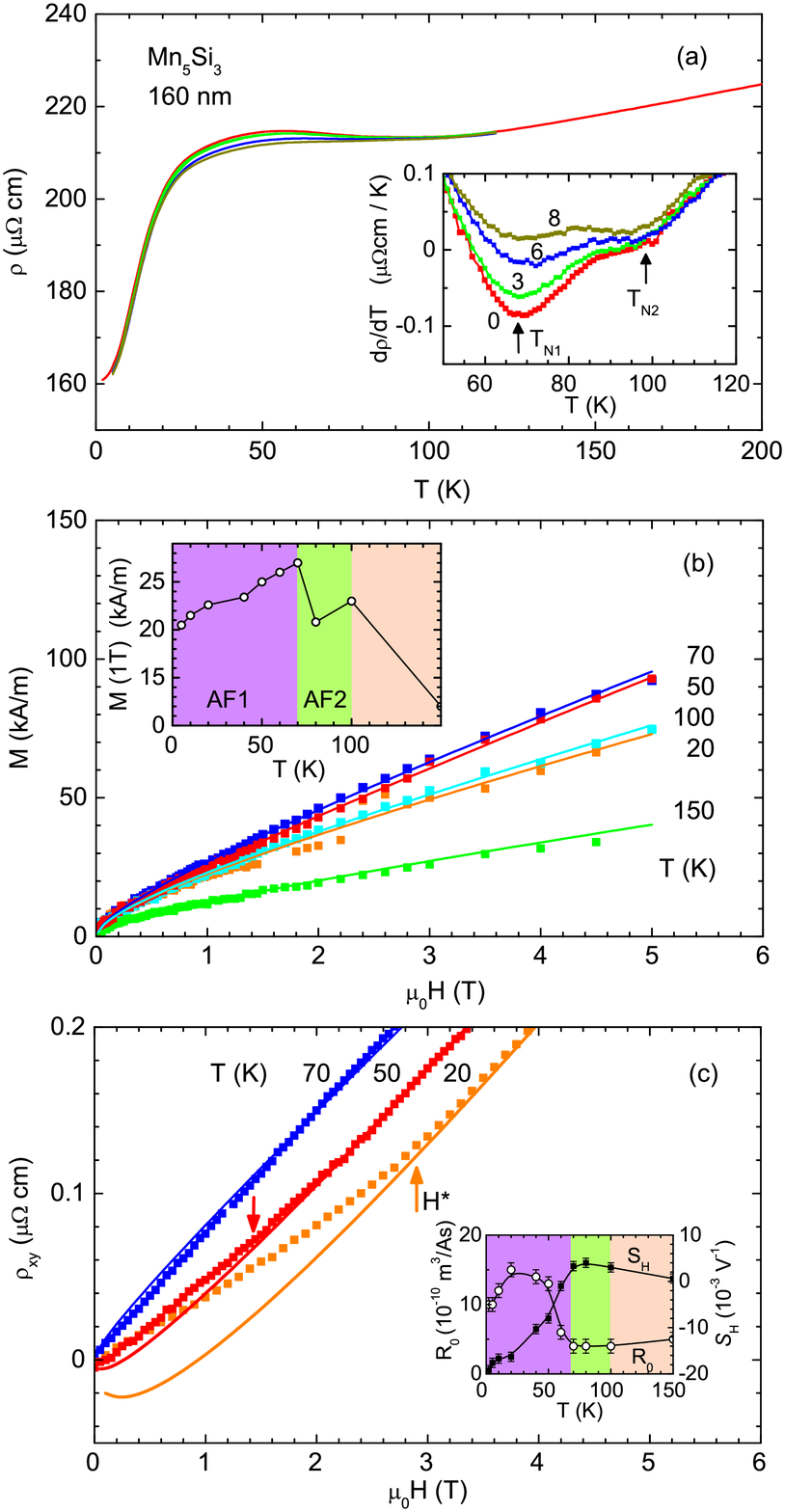}
\caption{\label{fig2}Mn$_5$Si$_3$ film (160 nm): (a) Resistivity $\rho_{xx}(T)$, inset shows the derivative of $\rho$. (b) Magnetization $M(H)$ for various temperatures $T$. Inset shows $M/H$ for $\mu_0H = 1$\,T. (c) Hall resistivity $\rho_{xy}(T)$ for various temperatures. Solid lines represent the calculated behavior of $\rho_{xy}$. Arrows indicate the critical field $H^{\ast}$ below which the data deviate from the calculated Hall resistivity, Eq. \ref{eq1}. Inset shows the temperature dependence of the Hall coefficients $R_0$ and $S_H$.}
\end{figure}

The longitudinal resistivity $\rho_{xx}$ [Fig. \ref{fig2}(a)] shows a behavior characteristic for an AFM with a linear temperature dependence in the paramagnetic regime and a 'hump-backed structure' \cite{meaden_conduction_1971} below the N\'{e}el temperature $T_{\rm N2}$ as observed earlier for polycrystals \cite{haug_electrical_1979}. From the derivative $d\rho/dT$ the transition temperatures $T_{\rm N2}$ = 99 K and $T_{\rm N1}$ = 68 K are obtained [Fig. \ref{fig2}(a), inset] in good agreement with values obtained for bulk polycrystals \cite{brown_low-temperature_1992,brown_antiferromagnetism_1995,gottschilch_study_2012}. $T_{\rm N2}$ is nearly independent of the magnetic field while the feature at $T_{\rm N1}$ diminishes with increasing field, see Fig. \ref{fig2}(a) inset. 
\\
The magnetization increases continously with field without saturation and with a slope that strongly changes with temperature, see Fig. \ref{fig2}(b). The inset shows that $M$(1 T) has a maximum close to $T_{\rm N1}$. Finally, Fig. \ref{fig2}(c) shows the Hall resistivity $\rho_{xy}$ for some selected temperatures in the noncollinear AF1 phase, at $T$\,= 50 K and 20 K well below $T_{\rm N1}$, and at $T = $ 70 K slightly above $T_{\rm N1}$. While at the latter $\rho_{xy}$ shows an almost linear behavior with magnetic field, the Hall resistivity in the noncollinear regime exhibits a kink at a field $H^{\ast}$ indicated by an arrow. 

For further analysis, we separated the contributions $R_0 B$ and $\rho_{xy}^{AH}$ to the Hall resistivity (Eq. \ref{eq1}) using the whole set of magnetotransport data with focus on the high-field regime and assuming $\rho_{xy}^{AH}=S_H\rho_{xx}^2M_z$ observed for ferromagnetic Mn$_5$Ge$_3$ (Ref. \onlinecite{zeng_linear_2006}). A possible additional contribution varying only linearly with $\rho_{xx}$ due to skew scattering does not change the obtained values $R_0$ and $S_H$ due to the high resistivity of the film \cite{surgers_large_2014} and is therefore not further taken into account. The inset in Fig. \ref{fig2}(c) shows that $R_0 \approx 5 \times 10^{-10} {\rm m}^3/{\rm As}$ and $S_H \approx 0$ above $T_{\rm N1}$ in the collinear antiferromagnetic and in the paramagnetic phases, i.e., the Hall effect is exclusively due to the ordinary contribution. Both Hall coefficients, $R_0$ and $S_H$ strongly change below $T_{\rm N1}$ indicating a considerable change of the Fermi surface. In particular, $S_H$ attains negative values of $-0.02\, {\rm V}^{^-1}$ at low temperatures. The solid lines in Fig. \ref{fig2}(c) show the field dependence of the Hall resistivity calculated from Eq. (1) with $R_0$ and $S_H$. The deviation between the data and the calculation (shaded area) is of the order of 20 n$\Omega$cm at $T$ = 20 K and is attributed to a THE arising from the noncollinear magnetic structure in the AF1 phase for magnetic fields below $H^{\ast}$ (Ref. \onlinecite{surgers_large_2014}).  

\subsection{Single crystal}
\begin{figure}
\includegraphics[width=0.9\columnwidth,clip=]{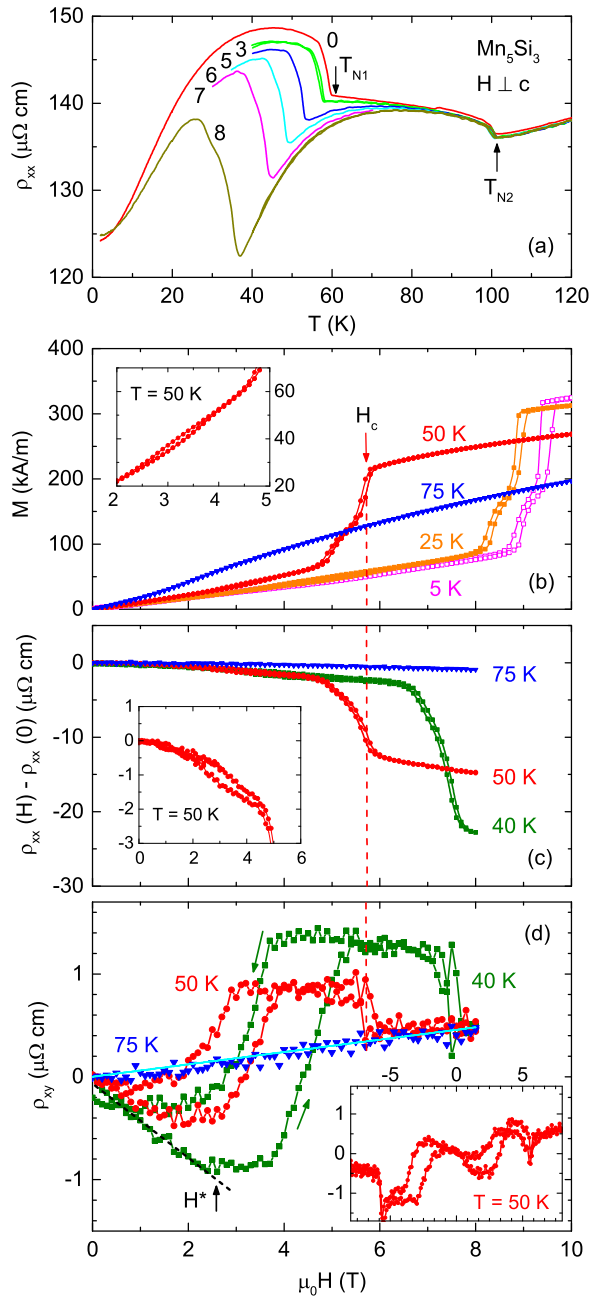}
\caption{\label{fig3}Mn$_5$Si$_3$ single crystal with $H$ applied along the \textit{z} direction perpendicularly to 
the crystallographic \textit{c} axis and current in the \textit{xy} plane. (a) Resistivity $\rho_{xx}(T)$, 
numbers indicate the applied magnetic field in T. (b) Magnetization $M(H)$ in the collinear ($T =$\, 75 K) and 
noncollinear regime ($T =$\, 50 K, 25 K, and 5 K). Inset shows a small hysteresis observed at $T$ = 50 K. 
(c) Magnetoresistivity. Inset shows a small hysteresis observed at $T$ = 50 K. (d) Hall resisitivity in the 
collinear ($T =$\, 75 K) and noncollinear regime ($T =$\, 40 and 50 K). Light blue line shows the ordinary Hall 
effect for $R_0 = 6 \times 10^{-10} {\rm m}^3 {\rm As}^{-1}$. Arrow and broken line indicate the critical field 
$H_{c}$ of the metamagnetic transition for $T$ = 50 K. Black dotted line indicates a linear decrease of $\rho_{xy}$ 
up to a field $H^{\ast}$ above which $\rho_{xy}$ increases to positive values. The inset shows raw data for positive and negative magnetic field at $T$ = 50 K.}
\end{figure}

The resistivity of the single crystal in the paramagnetic range above 100 K is reduced by only 35 \% with respect to the polycrystalline film, see Fig. \ref{fig3}(a). Yet, the features of $\rho_{xx}(T)$ at the two transitions $T_{\rm N2}$ = 100 K and $T_{\rm N1}$ = 60 K are much better resolved in the single crystal compared to the polycrystalline film. 
In the latter, averaging over all orientations in the electronic transport leads to a smearing of both transitions. Earlier resistivity measurements on Mn$_5$Si$_3$ single crystals also revealed two separate transitions \cite{vinokurova_magnetic_1990}. 
In a collinear AFM, the broad hump below the magnetic order-disorder N\'{e}el transition at $T_{\rm N2}$ is due to the formation of a magnetic superstructure larger than the crystal lattice and the new (smaller) Brillouin zone cuts the Fermi surface \cite{meaden_conduction_1971}. The clear separation of the two transitions in the resistivity not only persists in magnetic field but, in fact, becomes even more pronounced. While $T_{\rm N2}$ of the paramagnet/collinear-antiferromagnetic transition does not change with field, the transition temperature $T_{\rm N1}$ strongly decreases with increasing field. Neutron scattering and magnetic-susceptibility measurements on polycrystals already demonstrated that an applied magnetic field strongly reduces $T_{\rm N1}$ while $T_{\rm N2}$ remains unaffected \cite{silva_magnetic_2002,gottschilch_study_2012}. The field independence of $T_{\rm N2}$ is reminiscent of that of Mn$_3$Si (Ref. \onlinecite{pfleiderer_stability_2002}) and possibly due to the stability and strong anisotropy of Mn$_2$ moments. On the other hand, not only is $T_{N1}$ suppressed markedly with field but, in addition, the resistivity at $T_{N1}(H)$ decreases strongly. At $T_{N1}(8\, {\rm T}) = 37$\,K the resistivity even drops below the low-temperature residual resistivity $\rho_{xx}(T \rightarrow 0)$ of the noncollinear phase. This confirms that the spin-dependent scattering is much stronger in the noncollinear phase than in the collinear phase with a bipartite magnetic structure. 

The isothermal magnetization $M(H)$ plotted in Fig. \ref{fig3}(b) shows, for instance at $T$ = 50 K, a sudden increase from a low-field magnetization, increasing linearly from $M(0)$ = 0, to a much higher magnetization at a field $H_c$. Similar magnetization curves have been reported earlier \cite{al-kanani_magnetic_1995,vinokurova_magnetic_1990}. 
The jump at $H_c$ was attributed to a metamagnetic transition from a low-field noncollinear phase to a high-field phase akin to the collinear phase observed in zero field above $T_{\rm N1}$. The origin of a possible two-step transition at $H_c$ has to be investigated further. 
This metamagnetic transition is not observed in the 160-nm film possibly due to structural disorder and strong demagnetization effects, see Fig. \ref{fig2}(a). 
At $T$ = 5 K, $M$(10 T) = 310 kA/m perfectly agrees with previously published data \cite{vinokurova_magnetic_1990} 
and corresponds to a moment of 1.66 $\mu_{\rm B}$/Mn$_1$ or an average moment of 0.64 $\mu_{\rm B}$/Mn. We note that 
in the noncollinear phase at low fields the magnetization increases with increasing temperature while at fields above 
the metamagnetic transition the magnetization decreases with increasing temperature. The former resembles the thin-film 
behavior, cf. Fig. \ref{fig2}(b). A further detail is the weak hysteresis of $M(H)$ at $\approx$ 3 T well below the 
metamagnetic transition, exemplarily shown for $T$ = 50 K in Fig. \ref{fig3}(b) inset.
\\
The metamagnetic transition and the weak hysteresis are also observed in the magnetoresistivity [Fig. \ref{fig3}(c)] 
where the former leads to an almost 15 \% reduction of the resistivity at 40 K as reported earlier \cite{vinokurova_magnetic_1990}. 
In the collinear phase at 75 K, only a weak magnetoresistivity is observed.
\\
These two characteristics of the magnetic behavior of Mn$_5$Si$_3$ are most clearly observed in the Hall effect, exemplarily shown in Fig. \ref{fig3}(d) for three different temperatures. At 40 K, $\rho_{xy}$ roughly linearly decreases to negative values with increasing field until, at $H^{\ast} \approx$\, 2.5 T, $\rho_{xy}$ starts to increase with a sign change to positive values. Eventually, $\rho_{xy}$ saturates and becomes independent of $H$ until it sharply drops at a field corresponding to the metamagnetic transition at $H_{c}$. In the collinear phase at 75 K, $\rho_{xy}$ shows an only weak increase compatible with $S_H = 0$ and $R_0 = 6 \times 10^{-10} {\rm m}^3/{\rm As}$, similar to $R_0$ observed for the thin film above $T_{N1}$, cf. Fig. \ref{fig2}(c) inset. Moreover, for this orientation no hysteresis is observed for $\rho_{xy}$ when the field sweeps through zero, see the raw data for $T$ = 50 K in Fig. \ref{fig3}(d) (inset). Demagnetization effects have found to be negligible ($<$\,1\%) when taking into account $N_z$ = 0.635. Comparison of the AHE data confirms that at 40 and 50 K the (ordinary) Hall effect of the collinear phase at $T =$\, 75 K is recovered above $H_{c}$, suggesting that the field-induced phase below $T_{\rm N1}$ is similar to the zero-field collinear phase above $T_{\rm N1}$. 
Between $H^{\ast}$ and $H_{c}$ an additional field-induced magnetic phase exists with an AHE $\rho_{xy}$(3 T) = 0.9 $\mu \Omega$cm that is nearly independent of field and magnetization. At $T$ = 25 K an even larger $\rho_{xy}$(5 T) = 2 $\mu \Omega$cm corresponds to a Hall conductivity $\sigma_{xy} \approx \rho_{xy}/\rho_{xx}^2 = 102\, \Omega^{-1}{\rm cm}^{-1}$ despite a low magnetization of $M$(5 T) = 50 kA/m. Interestingly, $\rho_{xy}$ also seems to be independent of the magnetoresistivity $\rho_{xx}(H)$ in this field range, cf. Fig. \ref{fig3}(c). When reducing the field to zero, the field-induced intermediate phase becomes energetically unstable against the noncollinear arrangement which then gives rise to a strong decrease of $\rho_{xy}$ with a broad hysteresis of about 1 T.

\section{Discussion}
\begin{figure}
\includegraphics[width=0.7\columnwidth,clip=]{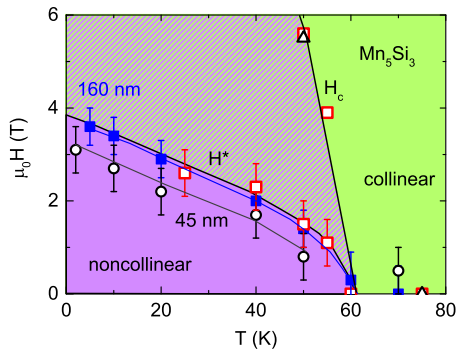}
\caption{\label{fig4}Magnetic fields $H^{\ast}$ and $H_c$ determined from the Hall effect for the single crystal (increasing field-sweep) with $H$ oriented perpendicularly to the crystallographic \textit{c} axis (square symbols), and $H^{\ast}$ of the 160-nm and 40-nm thick films \cite{surgers_large_2014}. Triangles indicate $H_c$ observed in $M(H)$.}
\end{figure}

In the paramagnetic and in the collinear antiferromagnetic phases above $T_{\rm N1}$ = 60 K, the Hall effect of both Mn$_5$Si$_3$ films and single crystal is dominated by the ordinary contribution $\propto R_0B$. At lower temperatures $T < T_{\rm N1}$, the AHE of the single crystal shows a different behavior compared to the 160-nm film. 
For the single crystal, it is difficult to separate the contributions arising from the ordinary Hall effect and AHE and we discuss the results only qualitatively. 
Fig. \ref{fig4} shows the characteristic fields $H^{\ast}$ for two films and the single crystal following a similar behavior. Below $H^{\ast}$, 
the noncollinear spin arrangement facilitates a THE. However, the magnitude of the THE is much more pronounced in the single crystal compared to the polycrystalline films. 
The different sign of the small THE signal in the latter requires further investiagations. The vanishing of the THE at $H^{\ast}$ suggests the formation of two different field-induced phases, one between $H^{\ast}$, the critical field of the noncollinear state, and $H_c$, and one above the field $H_c$ of the metamagnetic transition. 

Below $H^{\ast}$, the Mn$_1$ ordered moments disturb Mn$_2$ moments leading to a tilting of all moments and to noncollinearity \cite{brown_low-temperature_1992,silva_magnetic_2002,gottschilch_study_2012}. The magnetic configuration changes by an applied magnetic field due to a magnetostructural change from monoclinic back to orthorhombic symmetry with a decrease of the Mn$_1$-Mn$_1$ distance. This change causes a large inverse magnetocaloric effect \cite{gottschilch_study_2012}. Neutron scattering suggests that in 1 T the sample is already in the AF2 phase at $T = 63\, {\rm K}$ and likewise in 3.6 T at $T = 58\, {\rm K}$ (Ref. \onlinecite{silva_magnetic_2002}). In 4 T, the orthorhombic phase established at 50 K remains down to the lowest measuring temperature of $T$ = 5 K (Ref. \onlinecite{gottschilch_study_2012}). This interpretation of a stabilization of a collinear phase toward lower temperatures by a magnetic field is corroborated by the resistivity measurements [Fig. \ref{fig3}(a)]. However, the present results show that a collinear field-induced phase akin to the AF2 phase is only established above the metamagnetic transition $H_c$ where the Hall resisitivities above and below $T_{\rm N1}$ coincide. Between $H^{\ast}$ and $H_c$, an additional field-induced phase generating a large positive AHE with hysteresis is inferred from our Hall-effect data. In this regime, $\rho_{\rm xy}$ changes sign from negative to positive and becomes independent of $H$ and $M$. A possible scenario would be a weak ferromagnetic coupling of the Mn$_1$ moments still maintaining noncollinearity of the Mn$_1$ and Mn$_2$ moments.  

\section{Conclusion}
Comparison of the AHE of antiferromagnetic Mn$_5$Si$_3$ films and a single crystal provides deeper insight into the magnetotransport properties compared to magnetoresistivity. Below a characteristic field $H^{\ast}$, the noncollinear behavior of different Mn$_1$ and Mn$_2$ moments is maintained in the film as well as in the bulk crystal. This suggests that the large AHE attributed to noncollinearity is generated on a length scale of a few nm and is a local property not relying on long-range magnetic order. The variation of the AHE is more pronounced in the single crystal, for which an additional metamagnetic transition is observed at higher fields $H_{c}(T) > H^{\ast}(T)$. The metamagnetic transition is not observed in the film possibly due to structural and magnetic disorder and/or strong demagnetizing fields. First-principle caculations of the band structure of the noncollinear phase are strongly needed to obtain the Berry-phase curvature for a quantitative discussion of the results.

\begin{acknowledgments}
We thank P. Adelmann for structural characterization of the single crystal and A. Granovsky and F. Weber for helpful discussions.
\end{acknowledgments}


\begin{thebibliography}{38}%
\makeatletter
\providecommand \@ifxundefined [1]{%
 \@ifx{#1\undefined}
}%
\providecommand \@ifnum [1]{%
 \ifnum #1\expandafter \@firstoftwo
 \else \expandafter \@secondoftwo
 \fi
}%
\providecommand \@ifx [1]{%
 \ifx #1\expandafter \@firstoftwo
 \else \expandafter \@secondoftwo
 \fi
}%
\providecommand \natexlab [1]{#1}%
\providecommand \enquote  [1]{``#1''}%
\providecommand \bibnamefont  [1]{#1}%
\providecommand \bibfnamefont [1]{#1}%
\providecommand \citenamefont [1]{#1}%
\providecommand \href@noop [0]{\@secondoftwo}%
\providecommand \href [0]{\begingroup \@sanitize@url \@href}%
\providecommand \@href[1]{\@@startlink{#1}\@@href}%
\providecommand \@@href[1]{\endgroup#1\@@endlink}%
\providecommand \@sanitize@url [0]{\catcode `\\12\catcode `\$12\catcode
  `\&12\catcode `\#12\catcode `\^12\catcode `\_12\catcode `\%12\relax}%
\providecommand \@@startlink[1]{}%
\providecommand \@@endlink[0]{}%
\providecommand \url  [0]{\begingroup\@sanitize@url \@url }%
\providecommand \@url [1]{\endgroup\@href {#1}{\urlprefix }}%
\providecommand \urlprefix  [0]{URL }%
\providecommand \Eprint [0]{\href }%
\providecommand \doibase [0]{http://dx.doi.org/}%
\providecommand \selectlanguage [0]{\@gobble}%
\providecommand \bibinfo  [0]{\@secondoftwo}%
\providecommand \bibfield  [0]{\@secondoftwo}%
\providecommand \translation [1]{[#1]}%
\providecommand \BibitemOpen [0]{}%
\providecommand \bibitemStop [0]{}%
\providecommand \bibitemNoStop [0]{.\EOS\space}%
\providecommand \EOS [0]{\spacefactor3000\relax}%
\providecommand \BibitemShut  [1]{\csname bibitem#1\endcsname}%
\let\auto@bib@innerbib\@empty
\bibitem [{\citenamefont {MacDonald}\ and\ \citenamefont
  {Tsoi}(2011)}]{macdonald_antiferromagnetic_2011}%
  \BibitemOpen
  \bibfield  {author} {\bibinfo {author} {\bibfnamefont {A.~H.}\ \bibnamefont
  {MacDonald}}\ and\ \bibinfo {author} {\bibfnamefont {M.}~\bibnamefont
  {Tsoi}},\ }\href {\doibase 10.1098/rsta.2011.0014} {\bibfield  {journal}
  {\bibinfo  {journal} {Phil. Trans. Royal Soc. A}\ }\textbf {\bibinfo {volume}
  {369}},\ \bibinfo {pages} {3098} (\bibinfo {year} {2011})}\BibitemShut
  {NoStop}%
\bibitem [{\citenamefont {Cheng}\ \emph {et~al.}(2014)\citenamefont {Cheng},
  \citenamefont {Xiao}, \citenamefont {Niu},\ and\ \citenamefont
  {Brataas}}]{cheng_spin_2014}%
  \BibitemOpen
  \bibfield  {author} {\bibinfo {author} {\bibfnamefont {R.}~\bibnamefont
  {Cheng}}, \bibinfo {author} {\bibfnamefont {J.}~\bibnamefont {Xiao}},
  \bibinfo {author} {\bibfnamefont {Q.}~\bibnamefont {Niu}}, \ and\ \bibinfo
  {author} {\bibfnamefont {A.}~\bibnamefont {Brataas}},\ }\href {\doibase
  10.1103/PhysRevLett.113.057601} {\bibfield  {journal} {\bibinfo  {journal}
  {Phys. Rev. Lett.}\ }\textbf {\bibinfo {volume} {113}},\ \bibinfo {pages}
  {057601} (\bibinfo {year} {2014})}\BibitemShut {NoStop}%
\bibitem [{\citenamefont {Tveten}\ \emph {et~al.}(2013)\citenamefont {Tveten},
  \citenamefont {Qaiumzadeh}, \citenamefont {Tretiakov},\ and\ \citenamefont
  {Brataas}}]{tveten_staggered_2013}%
  \BibitemOpen
  \bibfield  {author} {\bibinfo {author} {\bibfnamefont {E.~G.}\ \bibnamefont
  {Tveten}}, \bibinfo {author} {\bibfnamefont {A.}~\bibnamefont {Qaiumzadeh}},
  \bibinfo {author} {\bibfnamefont {O.~A.}\ \bibnamefont {Tretiakov}}, \ and\
  \bibinfo {author} {\bibfnamefont {A.}~\bibnamefont {Brataas}},\ }\href
  {\doibase 10.1103/PhysRevLett.110.127208} {\bibfield  {journal} {\bibinfo
  {journal} {Phys. Rev. Lett.}\ }\textbf {\bibinfo {volume} {110}},\ \bibinfo
  {pages} {127208} (\bibinfo {year} {2013})}\BibitemShut {NoStop}%
\bibitem [{\citenamefont {Jungwirth}\ \emph {et~al.}(2015)\citenamefont
  {Jungwirth}, \citenamefont {Marti}, \citenamefont {Wadley},\ and\
  \citenamefont {Wunderlich}}]{jungwirth_antiferromagnetic_2015}%
  \BibitemOpen
  \bibfield  {author} {\bibinfo {author} {\bibfnamefont {T.}~\bibnamefont
  {Jungwirth}}, \bibinfo {author} {\bibfnamefont {X.}~\bibnamefont {Marti}},
  \bibinfo {author} {\bibfnamefont {P.}~\bibnamefont {Wadley}}, \ and\ \bibinfo
  {author} {\bibfnamefont {J.}~\bibnamefont {Wunderlich}},\ }\href
  {http://arxiv.org/abs/1509.05296} {\bibfield  {journal} {\bibinfo  {journal}
  {arXiv:1509.05296 [cond-mat]}\ } (\bibinfo {year} {2015})}\BibitemShut
  {NoStop}%
\bibitem [{\citenamefont {Kimel}\ \emph {et~al.}(2009)\citenamefont {Kimel},
  \citenamefont {Ivanov}, \citenamefont {Pisarev}, \citenamefont {Usachev},
  \citenamefont {Kirilyuk},\ and\ \citenamefont
  {Rasing}}]{kimel_inertia-driven_2009}%
  \BibitemOpen
  \bibfield  {author} {\bibinfo {author} {\bibfnamefont {A.~V.}\ \bibnamefont
  {Kimel}}, \bibinfo {author} {\bibfnamefont {B.~A.}\ \bibnamefont {Ivanov}},
  \bibinfo {author} {\bibfnamefont {R.~V.}\ \bibnamefont {Pisarev}}, \bibinfo
  {author} {\bibfnamefont {P.~A.}\ \bibnamefont {Usachev}}, \bibinfo {author}
  {\bibfnamefont {A.}~\bibnamefont {Kirilyuk}}, \ and\ \bibinfo {author}
  {\bibfnamefont {T.}~\bibnamefont {Rasing}},\ }\href {\doibase
  10.1038/nphys1369} {\bibfield  {journal} {\bibinfo  {journal} {Nat. Phys.}\
  }\textbf {\bibinfo {volume} {5}},\ \bibinfo {pages} {727} (\bibinfo {year}
  {2009})}\BibitemShut {NoStop}%
\bibitem [{\citenamefont {Nagaosa}\ \emph {et~al.}(2010)\citenamefont
  {Nagaosa}, \citenamefont {Sinova}, \citenamefont {Onoda}, \citenamefont
  {MacDonald},\ and\ \citenamefont {Ong}}]{nagaosa_anomalous_2010}%
  \BibitemOpen
  \bibfield  {author} {\bibinfo {author} {\bibfnamefont {N.}~\bibnamefont
  {Nagaosa}}, \bibinfo {author} {\bibfnamefont {J.}~\bibnamefont {Sinova}},
  \bibinfo {author} {\bibfnamefont {S.}~\bibnamefont {Onoda}}, \bibinfo
  {author} {\bibfnamefont {A.~H.}\ \bibnamefont {MacDonald}}, \ and\ \bibinfo
  {author} {\bibfnamefont {N.~P.}\ \bibnamefont {Ong}},\ }\href {\doibase
  10.1103/RevModPhys.82.1539} {\bibfield  {journal} {\bibinfo  {journal} {Rev.
  Mod. Phys.}\ }\textbf {\bibinfo {volume} {82}},\ \bibinfo {pages} {1539}
  (\bibinfo {year} {2010})}\BibitemShut {NoStop}%
\bibitem [{\citenamefont {Jungwirth}, \citenamefont {Niu},\ and\ \citenamefont
  {MacDonald}(2002)}]{jungwirth_anomalous_2002}%
  \BibitemOpen
  \bibfield  {author} {\bibinfo {author} {\bibfnamefont {T.}~\bibnamefont
  {Jungwirth}}, \bibinfo {author} {\bibfnamefont {Q.}~\bibnamefont {Niu}}, \
  and\ \bibinfo {author} {\bibfnamefont {A.~H.}\ \bibnamefont {MacDonald}},\
  }\href {\doibase 10.1103/PhysRevLett.88.207208} {\bibfield  {journal}
  {\bibinfo  {journal} {Phys. Rev. Lett.}\ }\textbf {\bibinfo {volume} {88}},\
  \bibinfo {pages} {207208} (\bibinfo {year} {2002})}\BibitemShut {NoStop}%
\bibitem [{\citenamefont {Miyasato}\ \emph {et~al.}(2007)\citenamefont
  {Miyasato}, \citenamefont {Abe}, \citenamefont {Fujii}, \citenamefont
  {Asamitsu}, \citenamefont {Onoda}, \citenamefont {Onose}, \citenamefont
  {Nagaosa},\ and\ \citenamefont {Tokura}}]{miyasato_crossover_2007}%
  \BibitemOpen
  \bibfield  {author} {\bibinfo {author} {\bibfnamefont {T.}~\bibnamefont
  {Miyasato}}, \bibinfo {author} {\bibfnamefont {N.}~\bibnamefont {Abe}},
  \bibinfo {author} {\bibfnamefont {T.}~\bibnamefont {Fujii}}, \bibinfo
  {author} {\bibfnamefont {A.}~\bibnamefont {Asamitsu}}, \bibinfo {author}
  {\bibfnamefont {S.}~\bibnamefont {Onoda}}, \bibinfo {author} {\bibfnamefont
  {Y.}~\bibnamefont {Onose}}, \bibinfo {author} {\bibfnamefont
  {N.}~\bibnamefont {Nagaosa}}, \ and\ \bibinfo {author} {\bibfnamefont
  {Y.}~\bibnamefont {Tokura}},\ }\href {\doibase 10.1103/PhysRevLett.99.086602}
  {\bibfield  {journal} {\bibinfo  {journal} {Phys. Rev. Lett.}\ }\textbf
  {\bibinfo {volume} {99}},\ \bibinfo {pages} {086602} (\bibinfo {year}
  {2007})}\BibitemShut {NoStop}%
\bibitem [{\citenamefont {Jiang}, \citenamefont {Zhou},\ and\ \citenamefont
  {Williams}(2010)}]{jiang_scaling_2010}%
  \BibitemOpen
  \bibfield  {author} {\bibinfo {author} {\bibfnamefont {W.}~\bibnamefont
  {Jiang}}, \bibinfo {author} {\bibfnamefont {X.~Z.}\ \bibnamefont {Zhou}}, \
  and\ \bibinfo {author} {\bibfnamefont {G.}~\bibnamefont {Williams}},\ }\href
  {\doibase 10.1103/PhysRevB.82.144424} {\bibfield  {journal} {\bibinfo
  {journal} {Phys. Rev. B}\ }\textbf {\bibinfo {volume} {82}},\ \bibinfo
  {pages} {144424} (\bibinfo {year} {2010})}\BibitemShut {NoStop}%
\bibitem [{\citenamefont {Turov}, \citenamefont {Shavrov},\ and\ \citenamefont
  {Irkhin}(1965)}]{turov_1965}%
  \BibitemOpen
  \bibfield  {author} {\bibinfo {author} {\bibfnamefont {E.~A.}\ \bibnamefont
  {Turov}}, \bibinfo {author} {\bibfnamefont {V.~G.}\ \bibnamefont {Shavrov}},
  \ and\ \bibinfo {author} {\bibfnamefont {Y.~P.}\ \bibnamefont {Irkhin}},\
  }\href@noop {} {\bibfield  {journal} {\bibinfo  {journal} {Soviet Physics
  JETP}\ }\textbf {\bibinfo {volume} {20}},\ \bibinfo {pages} {198} (\bibinfo
  {year} {1965})}\BibitemShut {NoStop}%
\bibitem [{\citenamefont {Taguchi}\ \emph {et~al.}(2001)\citenamefont
  {Taguchi}, \citenamefont {Oohara}, \citenamefont {Yoshizawa}, \citenamefont
  {Nagaosa},\ and\ \citenamefont {Tokura}}]{taguchi_spin_2001}%
  \BibitemOpen
  \bibfield  {author} {\bibinfo {author} {\bibfnamefont {Y.}~\bibnamefont
  {Taguchi}}, \bibinfo {author} {\bibfnamefont {Y.}~\bibnamefont {Oohara}},
  \bibinfo {author} {\bibfnamefont {H.}~\bibnamefont {Yoshizawa}}, \bibinfo
  {author} {\bibfnamefont {N.}~\bibnamefont {Nagaosa}}, \ and\ \bibinfo
  {author} {\bibfnamefont {Y.}~\bibnamefont {Tokura}},\ }\href {\doibase
  10.1126/science.1058161} {\bibfield  {journal} {\bibinfo  {journal}
  {Science}\ }\textbf {\bibinfo {volume} {291}},\ \bibinfo {pages} {2573}
  (\bibinfo {year} {2001})}\BibitemShut {NoStop}%
\bibitem [{\citenamefont {Machida}\ \emph {et~al.}(2010)\citenamefont
  {Machida}, \citenamefont {Nakatsuji}, \citenamefont {Onoda}, \citenamefont
  {Tayama},\ and\ \citenamefont {Sakakibara}}]{machida_time-reversal_2010}%
  \BibitemOpen
  \bibfield  {author} {\bibinfo {author} {\bibfnamefont {Y.}~\bibnamefont
  {Machida}}, \bibinfo {author} {\bibfnamefont {S.}~\bibnamefont {Nakatsuji}},
  \bibinfo {author} {\bibfnamefont {S.}~\bibnamefont {Onoda}}, \bibinfo
  {author} {\bibfnamefont {T.}~\bibnamefont {Tayama}}, \ and\ \bibinfo {author}
  {\bibfnamefont {T.}~\bibnamefont {Sakakibara}},\ }\href {\doibase
  10.1038/nature08680} {\bibfield  {journal} {\bibinfo  {journal} {Nature}\
  }\textbf {\bibinfo {volume} {463}},\ \bibinfo {pages} {210} (\bibinfo {year}
  {2010})}\BibitemShut {NoStop}%
\bibitem [{\citenamefont {Ueland}\ \emph {et~al.}(2012)\citenamefont {Ueland},
  \citenamefont {Miclea}, \citenamefont {Kato}, \citenamefont
  {Ayala-Valenzuela}, \citenamefont {McDonald}, \citenamefont {Okazaki},
  \citenamefont {Tobash}, \citenamefont {Torrez}, \citenamefont {Ronning},
  \citenamefont {Movshovich}, \citenamefont {Fisk}, \citenamefont {Bauer},
  \citenamefont {Martin},\ and\ \citenamefont
  {Thompson}}]{ueland_controllable_2012}%
  \BibitemOpen
  \bibfield  {author} {\bibinfo {author} {\bibfnamefont {B.~G.}\ \bibnamefont
  {Ueland}}, \bibinfo {author} {\bibfnamefont {C.~F.}\ \bibnamefont {Miclea}},
  \bibinfo {author} {\bibfnamefont {Y.}~\bibnamefont {Kato}}, \bibinfo {author}
  {\bibfnamefont {O.}~\bibnamefont {Ayala-Valenzuela}}, \bibinfo {author}
  {\bibfnamefont {R.~D.}\ \bibnamefont {McDonald}}, \bibinfo {author}
  {\bibfnamefont {R.}~\bibnamefont {Okazaki}}, \bibinfo {author} {\bibfnamefont
  {P.~H.}\ \bibnamefont {Tobash}}, \bibinfo {author} {\bibfnamefont {M.~A.}\
  \bibnamefont {Torrez}}, \bibinfo {author} {\bibfnamefont {F.}~\bibnamefont
  {Ronning}}, \bibinfo {author} {\bibfnamefont {R.}~\bibnamefont {Movshovich}},
  \bibinfo {author} {\bibfnamefont {Z.}~\bibnamefont {Fisk}}, \bibinfo {author}
  {\bibfnamefont {E.~D.}\ \bibnamefont {Bauer}}, \bibinfo {author}
  {\bibfnamefont {I.}~\bibnamefont {Martin}}, \ and\ \bibinfo {author}
  {\bibfnamefont {J.~D.}\ \bibnamefont {Thompson}},\ }\href {\doibase
  10.1038/ncomms2075} {\bibfield  {journal} {\bibinfo  {journal} {Nat.
  Commun.}\ }\textbf {\bibinfo {volume} {3}},\ \bibinfo {pages} {1067}
  (\bibinfo {year} {2012})}\BibitemShut {NoStop}%
\bibitem [{\citenamefont {Nagaosa}\ and\ \citenamefont
  {Tokura}(2013)}]{nagaosa_topological_2013}%
  \BibitemOpen
  \bibfield  {author} {\bibinfo {author} {\bibfnamefont {N.}~\bibnamefont
  {Nagaosa}}\ and\ \bibinfo {author} {\bibfnamefont {Y.}~\bibnamefont
  {Tokura}},\ }\href {\doibase 10.1038/nnano.2013.243} {\bibfield  {journal}
  {\bibinfo  {journal} {Nat. Nanotechn.}\ }\textbf {\bibinfo {volume} {8}},\
  \bibinfo {pages} {899} (\bibinfo {year} {2013})}\BibitemShut {NoStop}%
\bibitem [{\citenamefont {Schulz}\ \emph {et~al.}(2012)\citenamefont {Schulz},
  \citenamefont {Ritz}, \citenamefont {Bauer}, \citenamefont {Halder},
  \citenamefont {Wagner}, \citenamefont {Franz}, \citenamefont {Pfleiderer},
  \citenamefont {Everschor}, \citenamefont {Garst},\ and\ \citenamefont
  {Rosch}}]{schulz_emergent_2012}%
  \BibitemOpen
  \bibfield  {author} {\bibinfo {author} {\bibfnamefont {T.}~\bibnamefont
  {Schulz}}, \bibinfo {author} {\bibfnamefont {R.}~\bibnamefont {Ritz}},
  \bibinfo {author} {\bibfnamefont {A.}~\bibnamefont {Bauer}}, \bibinfo
  {author} {\bibfnamefont {M.}~\bibnamefont {Halder}}, \bibinfo {author}
  {\bibfnamefont {M.}~\bibnamefont {Wagner}}, \bibinfo {author} {\bibfnamefont
  {C.}~\bibnamefont {Franz}}, \bibinfo {author} {\bibfnamefont
  {C.}~\bibnamefont {Pfleiderer}}, \bibinfo {author} {\bibfnamefont
  {K.}~\bibnamefont {Everschor}}, \bibinfo {author} {\bibfnamefont
  {M.}~\bibnamefont {Garst}}, \ and\ \bibinfo {author} {\bibfnamefont
  {A.}~\bibnamefont {Rosch}},\ }\href {\doibase 10.1038/nphys2231} {\bibfield
  {journal} {\bibinfo  {journal} {Nat. Phys.}\ }\textbf {\bibinfo {volume}
  {8}},\ \bibinfo {pages} {301} (\bibinfo {year} {2012})}\BibitemShut {NoStop}%
\bibitem [{\citenamefont {Hoffmann}\ \emph {et~al.}(2015)\citenamefont
  {Hoffmann}, \citenamefont {Weischenberg}, \citenamefont {Dupé},
  \citenamefont {Freimuth}, \citenamefont {Ferriani}, \citenamefont
  {Mokrousov},\ and\ \citenamefont {Heinze}}]{hoffmann_topological_2015}%
  \BibitemOpen
  \bibfield  {author} {\bibinfo {author} {\bibfnamefont {M.}~\bibnamefont
  {Hoffmann}}, \bibinfo {author} {\bibfnamefont {J.}~\bibnamefont
  {Weischenberg}}, \bibinfo {author} {\bibfnamefont {B.}~\bibnamefont {Dupé}},
  \bibinfo {author} {\bibfnamefont {F.}~\bibnamefont {Freimuth}}, \bibinfo
  {author} {\bibfnamefont {P.}~\bibnamefont {Ferriani}}, \bibinfo {author}
  {\bibfnamefont {Y.}~\bibnamefont {Mokrousov}}, \ and\ \bibinfo {author}
  {\bibfnamefont {S.}~\bibnamefont {Heinze}},\ }\href {\doibase
  10.1103/PhysRevB.92.020401} {\bibfield  {journal} {\bibinfo  {journal} {Phys.
  Rev. B}\ }\textbf {\bibinfo {volume} {92}},\ \bibinfo {pages} {020401}
  (\bibinfo {year} {2015})}\BibitemShut {NoStop}%
\bibitem [{\citenamefont {Chen}, \citenamefont {Niu},\ and\ \citenamefont
  {MacDonald}(2014)}]{chen_anomalous_2014}%
  \BibitemOpen
  \bibfield  {author} {\bibinfo {author} {\bibfnamefont {H.}~\bibnamefont
  {Chen}}, \bibinfo {author} {\bibfnamefont {Q.}~\bibnamefont {Niu}}, \ and\
  \bibinfo {author} {\bibfnamefont {A.}~\bibnamefont {MacDonald}},\ }\href
  {\doibase 10.1103/PhysRevLett.112.017205} {\bibfield  {journal} {\bibinfo
  {journal} {Phys. Rev. Lett.}\ }\textbf {\bibinfo {volume} {112}},\ \bibinfo
  {pages} {017205} (\bibinfo {year} {2014})}\BibitemShut {NoStop}%
\bibitem [{\citenamefont {K\"ubler}\ and\ \citenamefont
  {Felser}(2014)}]{kubler_non-collinear_2014}%
  \BibitemOpen
  \bibfield  {author} {\bibinfo {author} {\bibfnamefont {J.}~\bibnamefont
  {K\"ubler}}\ and\ \bibinfo {author} {\bibfnamefont {C.}~\bibnamefont
  {Felser}},\ }\href {\doibase 10.1209/0295-5075/108/67001} {\bibfield
  {journal} {\bibinfo  {journal} {EPL}\ }\textbf {\bibinfo {volume} {108}},\
  \bibinfo {pages} {67001} (\bibinfo {year} {2014})}\BibitemShut {NoStop}%
\bibitem [{\citenamefont {Gomonay}(2015)}]{gomonay_berry-phase_2015}%
  \BibitemOpen
  \bibfield  {author} {\bibinfo {author} {\bibfnamefont {O.}~\bibnamefont
  {Gomonay}},\ }\href {\doibase 10.1103/PhysRevB.91.144421} {\bibfield
  {journal} {\bibinfo  {journal} {Phys. Rev. B}\ }\textbf {\bibinfo {volume}
  {91}},\ \bibinfo {pages} {144421} (\bibinfo {year} {2015})}\BibitemShut
  {NoStop}%
\bibitem [{\citenamefont {S\"urgers}\ \emph {et~al.}(2014)\citenamefont
  {S\"urgers}, \citenamefont {Fischer}, \citenamefont {Winkel},\ and\
  \citenamefont {L\"ohneysen}}]{surgers_large_2014}%
  \BibitemOpen
  \bibfield  {author} {\bibinfo {author} {\bibfnamefont {C.}~\bibnamefont
  {S\"urgers}}, \bibinfo {author} {\bibfnamefont {G.}~\bibnamefont {Fischer}},
  \bibinfo {author} {\bibfnamefont {P.}~\bibnamefont {Winkel}}, \ and\ \bibinfo
  {author} {\bibfnamefont {H.~v.}\ \bibnamefont {L\"ohneysen}},\ }\href
  {\doibase 10.1038/ncomms4400} {\bibfield  {journal} {\bibinfo  {journal}
  {Nat. Commun.}\ }\textbf {\bibinfo {volume} {5}},\ \bibinfo {pages} {3400}
  (\bibinfo {year} {2014})}\BibitemShut {NoStop}%
\bibitem [{\citenamefont {Stroppa}\ and\ \citenamefont
  {Peressi}(2007)}]{stroppa_competing_2007}%
  \BibitemOpen
  \bibfield  {author} {\bibinfo {author} {\bibfnamefont {A.}~\bibnamefont
  {Stroppa}}\ and\ \bibinfo {author} {\bibfnamefont {M.}~\bibnamefont
  {Peressi}},\ }\href {\doibase 10.1002/pssa.200673014} {\bibfield  {journal}
  {\bibinfo  {journal} {phys. stat. sol. (a)}\ }\textbf {\bibinfo {volume}
  {204}},\ \bibinfo {pages} {44–52} (\bibinfo {year}
  {2007})}\BibitemShut {NoStop}%
\bibitem [{\citenamefont {Zeng}\ \emph {et~al.}(2003)\citenamefont {Zeng},
  \citenamefont {Erwin}, \citenamefont {Feldman}, \citenamefont {Li},
  \citenamefont {Jin}, \citenamefont {Song}, \citenamefont {Thompson},\ and\
  \citenamefont {Weitering}}]{zeng_epitaxial_2003}%
  \BibitemOpen
  \bibfield  {author} {\bibinfo {author} {\bibfnamefont {C.}~\bibnamefont
  {Zeng}}, \bibinfo {author} {\bibfnamefont {S.~C.}\ \bibnamefont {Erwin}},
  \bibinfo {author} {\bibfnamefont {L.~C.}\ \bibnamefont {Feldman}}, \bibinfo
  {author} {\bibfnamefont {A.~P.}\ \bibnamefont {Li}}, \bibinfo {author}
  {\bibfnamefont {R.}~\bibnamefont {Jin}}, \bibinfo {author} {\bibfnamefont
  {Y.}~\bibnamefont {Song}}, \bibinfo {author} {\bibfnamefont {J.~R.}\
  \bibnamefont {Thompson}}, \ and\ \bibinfo {author} {\bibfnamefont {H.~H.}\
  \bibnamefont {Weitering}},\ }\href {\doibase doi:10.1063/1.1633684}
  {\bibfield  {journal} {\bibinfo  {journal} {Appl. Phys. Lett.}\ }\textbf
  {\bibinfo {volume} {83}},\ \bibinfo {pages} {5002} (\bibinfo {year}
  {2003})}\BibitemShut {NoStop}%
\bibitem [{\citenamefont {S\"urgers}\ \emph {et~al.}(2008)\citenamefont
  {S\"urgers}, \citenamefont {Potzger}, \citenamefont {Strache}, \citenamefont
  {M\"uller}, \citenamefont {Fischer}, \citenamefont {Joshi},\ and\
  \citenamefont {v.~L\"ohneysen}}]{surgers_magnetic_2008}%
  \BibitemOpen
  \bibfield  {author} {\bibinfo {author} {\bibfnamefont {C.}~\bibnamefont
  {S\"urgers}}, \bibinfo {author} {\bibfnamefont {K.}~\bibnamefont {Potzger}},
  \bibinfo {author} {\bibfnamefont {T.}~\bibnamefont {Strache}}, \bibinfo
  {author} {\bibfnamefont {W.}~\bibnamefont {M\"uller}}, \bibinfo {author}
  {\bibfnamefont {G.}~\bibnamefont {Fischer}}, \bibinfo {author} {\bibfnamefont
  {N.}~\bibnamefont {Joshi}}, \ and\ \bibinfo {author} {\bibfnamefont
  {H.}~\bibnamefont {v.~L\"ohneysen}},\ }\href {\doibase doi:10.1063/1.2969403}
  {\bibfield  {journal} {\bibinfo  {journal} {Appl. Phys. Lett.}\ }\textbf
  {\bibinfo {volume} {93}},\ \bibinfo {pages} {062503} (\bibinfo {year}
  {2008})}\BibitemShut {NoStop}%
\bibitem [{\citenamefont {Slipukhina}\ \emph {et~al.}(2009)\citenamefont
  {Slipukhina}, \citenamefont {Arras}, \citenamefont {Mavropoulos},\ and\
  \citenamefont {Pochet}}]{slipukhina_simulation_2009}%
  \BibitemOpen
  \bibfield  {author} {\bibinfo {author} {\bibfnamefont {I.}~\bibnamefont
  {Slipukhina}}, \bibinfo {author} {\bibfnamefont {E.}~\bibnamefont {Arras}},
  \bibinfo {author} {\bibfnamefont {P.}~\bibnamefont {Mavropoulos}}, \ and\
  \bibinfo {author} {\bibfnamefont {P.}~\bibnamefont {Pochet}},\ }\href
  {\doibase doi:10.1063/1.3134482} {\bibfield  {journal} {\bibinfo  {journal}
  {Appl. Phys. Lett.}\ }\textbf {\bibinfo {volume} {94}},\ \bibinfo {pages}
  {192505} (\bibinfo {year} {2009})}\BibitemShut {NoStop}%
\bibitem [{\citenamefont {Thanh}\ \emph {et~al.}(2013)\citenamefont {Thanh},
  \citenamefont {Spiesser}, \citenamefont {Dau}, \citenamefont {Olive-Mendez},
  \citenamefont {Michez},\ and\ \citenamefont {Petit}}]{thanh_epitaxial_2013}%
  \BibitemOpen
  \bibfield  {author} {\bibinfo {author} {\bibfnamefont {V.~L.}\ \bibnamefont
  {Thanh}}, \bibinfo {author} {\bibfnamefont {A.}~\bibnamefont {Spiesser}},
  \bibinfo {author} {\bibfnamefont {M.-T.}\ \bibnamefont {Dau}}, \bibinfo
  {author} {\bibfnamefont {S.~F.}\ \bibnamefont {Olive-Mendez}}, \bibinfo
  {author} {\bibfnamefont {L.~A.}\ \bibnamefont {Michez}}, \ and\ \bibinfo
  {author} {\bibfnamefont {M.}~\bibnamefont {Petit}},\ }\href {\doibase
  10.1088/2043-6262/4/4/043002} {\bibfield  {journal} {\bibinfo  {journal}
  {Adv. Nat. Sci: Nanosci. Nanotechnol.}\ }\textbf {\bibinfo {volume} {4}},\
  \bibinfo {pages} {043002} (\bibinfo {year} {2013})}\BibitemShut {NoStop}%
\bibitem [{\citenamefont {Fischer}\ \emph {et~al.}(2014)\citenamefont
  {Fischer}, \citenamefont {Chang}, \citenamefont {S\"urgers}, \citenamefont
  {Rolseth}, \citenamefont {Reiter}, \citenamefont {Stefanov}, \citenamefont
  {Chiussi}, \citenamefont {Tang}, \citenamefont {Wang},\ and\ \citenamefont
  {Schulze}}]{fischer_hanle-effect_2014}%
  \BibitemOpen
  \bibfield  {author} {\bibinfo {author} {\bibfnamefont {I.~A.}\ \bibnamefont
  {Fischer}}, \bibinfo {author} {\bibfnamefont {L.-T.}\ \bibnamefont {Chang}},
  \bibinfo {author} {\bibfnamefont {C.}~\bibnamefont {S\"urgers}}, \bibinfo
  {author} {\bibfnamefont {E.}~\bibnamefont {Rolseth}}, \bibinfo {author}
  {\bibfnamefont {S.}~\bibnamefont {Reiter}}, \bibinfo {author} {\bibfnamefont
  {S.}~\bibnamefont {Stefanov}}, \bibinfo {author} {\bibfnamefont
  {S.}~\bibnamefont {Chiussi}}, \bibinfo {author} {\bibfnamefont
  {J.}~\bibnamefont {Tang}}, \bibinfo {author} {\bibfnamefont {K.~L.}\
  \bibnamefont {Wang}}, \ and\ \bibinfo {author} {\bibfnamefont
  {J.}~\bibnamefont {Schulze}},\ }\href {\doibase 10.1063/1.4903233} {\bibfield
   {journal} {\bibinfo  {journal} {Appl. Phys. Lett.}\ }\textbf {\bibinfo
  {volume} {105}},\ \bibinfo {pages} {222408} (\bibinfo {year}
  {2014})}\BibitemShut {NoStop}%
\bibitem [{\citenamefont {Forsyth}\ and\ \citenamefont
  {Brown}(1990)}]{forsyth_spatial_1990}%
  \BibitemOpen
  \bibfield  {author} {\bibinfo {author} {\bibfnamefont {J.~B.}\ \bibnamefont
  {Forsyth}}\ and\ \bibinfo {author} {\bibfnamefont {P.~J.}\ \bibnamefont
  {Brown}},\ }\href {\doibase 10.1088/0953-8984/2/11/014} {\bibfield  {journal}
  {\bibinfo  {journal} {J. Phys.: Condens. Matter}\ }\textbf {\bibinfo {volume}
  {2}},\ \bibinfo {pages} {2713} (\bibinfo {year} {1990})}\BibitemShut
  {NoStop}%
\bibitem [{\citenamefont {Gottschilch}\ \emph {et~al.}(2012)\citenamefont
  {Gottschilch}, \citenamefont {Gourdon}, \citenamefont {Persson},
  \citenamefont {de~la Cruz}, \citenamefont {Petricek},\ and\ \citenamefont
  {Brueckel}}]{gottschilch_study_2012}%
  \BibitemOpen
  \bibfield  {author} {\bibinfo {author} {\bibfnamefont {M.}~\bibnamefont
  {Gottschilch}}, \bibinfo {author} {\bibfnamefont {O.}~\bibnamefont
  {Gourdon}}, \bibinfo {author} {\bibfnamefont {J.}~\bibnamefont {Persson}},
  \bibinfo {author} {\bibfnamefont {C.}~\bibnamefont {de~la Cruz}}, \bibinfo
  {author} {\bibfnamefont {V.}~\bibnamefont {Petricek}}, \ and\ \bibinfo
  {author} {\bibfnamefont {T.}~\bibnamefont {Brueckel}},\ }\href {\doibase
  10.1039/C2JM00154C} {\bibfield  {journal} {\bibinfo  {journal} {J. Mater.
  Chem.}\ }\textbf {\bibinfo {volume} {22}},\ \bibinfo {pages} {15275}
  (\bibinfo {year} {2012})}\BibitemShut {NoStop}%
\bibitem [{\citenamefont {Brown}\ \emph {et~al.}(1992)\citenamefont {Brown},
  \citenamefont {Forsyth}, \citenamefont {Nunez},\ and\ \citenamefont
  {Tasset}}]{brown_low-temperature_1992}%
  \BibitemOpen
  \bibfield  {author} {\bibinfo {author} {\bibfnamefont {P.~J.}\ \bibnamefont
  {Brown}}, \bibinfo {author} {\bibfnamefont {J.~B.}\ \bibnamefont {Forsyth}},
  \bibinfo {author} {\bibfnamefont {V.}~\bibnamefont {Nunez}}, \ and\ \bibinfo
  {author} {\bibfnamefont {F.}~\bibnamefont {Tasset}},\ }\href {\doibase
  10.1088/0953-8984/4/49/029} {\bibfield  {journal} {\bibinfo  {journal} {J.
  Phys.: Condens. Matter}\ }\textbf {\bibinfo {volume} {4}},\ \bibinfo {pages}
  {10025} (\bibinfo {year} {1992})}\BibitemShut {NoStop}%
\bibitem [{\citenamefont {Brown}\ and\ \citenamefont
  {Forsyth}(1995)}]{brown_antiferromagnetism_1995}%
  \BibitemOpen
  \bibfield  {author} {\bibinfo {author} {\bibfnamefont {P.~J.}\ \bibnamefont
  {Brown}}\ and\ \bibinfo {author} {\bibfnamefont {J.~B.}\ \bibnamefont
  {Forsyth}},\ }\href {\doibase 10.1088/0953-8984/7/39/004} {\bibfield
  {journal} {\bibinfo  {journal} {J. Phys.: Condens. Matter}\ }\textbf
  {\bibinfo {volume} {7}},\ \bibinfo {pages} {7619} (\bibinfo {year}
  {1995})}\BibitemShut {NoStop}%
\bibitem [{\citenamefont {Silva}, \citenamefont {Brown},\ and\ \citenamefont
  {Forsyth}(2002)}]{silva_magnetic_2002}%
  \BibitemOpen
  \bibfield  {author} {\bibinfo {author} {\bibfnamefont {M.~R.}\ \bibnamefont
  {Silva}}, \bibinfo {author} {\bibfnamefont {P.~J.}\ \bibnamefont {Brown}}, \
  and\ \bibinfo {author} {\bibfnamefont {J.~B.}\ \bibnamefont {Forsyth}},\
  }\href {\doibase 10.1088/0953-8984/14/37/307} {\bibfield  {journal} {\bibinfo
   {journal} {J. Phys.: Condens. Matter}\ }\textbf {\bibinfo {volume} {14}},\
  \bibinfo {pages} {8707} (\bibinfo {year} {2002})}\BibitemShut {NoStop}%
\bibitem [{\citenamefont {Gopalakrishnan}\ \emph {et~al.}(2008)\citenamefont
  {Gopalakrishnan}, \citenamefont {S\"urgers}, \citenamefont {Montbrun},
  \citenamefont {Singh}, \citenamefont {Uhlarz},\ and\ \citenamefont
  {L\"ohneysen}}]{gopalakrishnan_electronic_2008}%
  \BibitemOpen
  \bibfield  {author} {\bibinfo {author} {\bibfnamefont {B.}~\bibnamefont
  {Gopalakrishnan}}, \bibinfo {author} {\bibfnamefont {C.}~\bibnamefont
  {S\"urgers}}, \bibinfo {author} {\bibfnamefont {R.}~\bibnamefont {Montbrun}},
  \bibinfo {author} {\bibfnamefont {A.}~\bibnamefont {Singh}}, \bibinfo
  {author} {\bibfnamefont {M.}~\bibnamefont {Uhlarz}}, \ and\ \bibinfo {author}
  {\bibfnamefont {H.~v.}\ \bibnamefont {L\"ohneysen}},\ }\href {\doibase
  10.1103/PhysRevB.77.104414} {\bibfield  {journal} {\bibinfo  {journal} {Phys.
  Rev. B}\ }\textbf {\bibinfo {volume} {77}},\ \bibinfo {pages} {104414}
  (\bibinfo {year} {2008})}\BibitemShut {NoStop}%
\bibitem [{\citenamefont {Meaden}(1971)}]{meaden_conduction_1971}%
  \BibitemOpen
  \bibfield  {author} {\bibinfo {author} {\bibfnamefont {G.~T.}\ \bibnamefont
  {Meaden}},\ }\href {\doibase 10.1080/00107517108205267} {\bibfield  {journal}
  {\bibinfo  {journal} {Contemp. Phys.}\ }\textbf {\bibinfo {volume} {12}},\
  \bibinfo {pages} {313} (\bibinfo {year} {1971})}\BibitemShut {NoStop}%
\bibitem [{\citenamefont {Haug}, \citenamefont {Kappel},\ and\ \citenamefont
  {Ja\'{e}gle}(1979)}]{haug_electrical_1979}%
  \BibitemOpen
  \bibfield  {author} {\bibinfo {author} {\bibfnamefont {R.}~\bibnamefont
  {Haug}}, \bibinfo {author} {\bibfnamefont {G.}~\bibnamefont {Kappel}}, \ and\
  \bibinfo {author} {\bibfnamefont {A.}~\bibnamefont {Ja\'{e}gle}},\ }\href
  {\doibase 10.1002/pssa.2210550132} {\bibfield  {journal} {\bibinfo  {journal}
  {phys. stat. sol. (a)}\ }\textbf {\bibinfo {volume} {55}},\ \bibinfo {pages}
  {285} (\bibinfo {year} {1979})}\BibitemShut {NoStop}%
\bibitem [{\citenamefont {Zeng}\ \emph {et~al.}(2006)\citenamefont {Zeng},
  \citenamefont {Yao}, \citenamefont {Niu},\ and\ \citenamefont
  {Weitering}}]{zeng_linear_2006}%
  \BibitemOpen
  \bibfield  {author} {\bibinfo {author} {\bibfnamefont {C.}~\bibnamefont
  {Zeng}}, \bibinfo {author} {\bibfnamefont {Y.}~\bibnamefont {Yao}}, \bibinfo
  {author} {\bibfnamefont {Q.}~\bibnamefont {Niu}}, \ and\ \bibinfo {author}
  {\bibfnamefont {H.~H.}\ \bibnamefont {Weitering}},\ }\href {\doibase
  10.1103/PhysRevLett.96.037204} {\bibfield  {journal} {\bibinfo  {journal}
  {Phys. Rev. Lett.}\ }\textbf {\bibinfo {volume} {96}},\ \bibinfo {pages}
  {037204} (\bibinfo {year} {2006})}\BibitemShut {NoStop}%
\bibitem [{\citenamefont {Vinokurova}\ \emph {et~al.}(1990)\citenamefont
  {Vinokurova}, \citenamefont {Ivanov}, \citenamefont {Kulatov},\ and\
  \citenamefont {Vlasov}}]{vinokurova_magnetic_1990}%
  \BibitemOpen
  \bibfield  {author} {\bibinfo {author} {\bibfnamefont {L.}~\bibnamefont
  {Vinokurova}}, \bibinfo {author} {\bibfnamefont {V.}~\bibnamefont {Ivanov}},
  \bibinfo {author} {\bibfnamefont {E.}~\bibnamefont {Kulatov}}, \ and\
  \bibinfo {author} {\bibfnamefont {A.}~\bibnamefont {Vlasov}},\ }\href
  {\doibase 10.1016/S0304-8853(10)80040-X} {\bibfield  {journal} {\bibinfo
  {journal} {Journ. Magn. Magn. Mater.}\ }\textbf {\bibinfo {volume} {90-91}},\
  \bibinfo {pages} {121} (\bibinfo {year} {1990})}\BibitemShut {NoStop}%
\bibitem [{\citenamefont {Pfleiderer}, \citenamefont {B{\oe}uf},\ and\
  \citenamefont {L\"ohneysen}(2002)}]{pfleiderer_stability_2002}%
  \BibitemOpen
  \bibfield  {author} {\bibinfo {author} {\bibfnamefont {C.}~\bibnamefont
  {Pfleiderer}}, \bibinfo {author} {\bibfnamefont {J.}~\bibnamefont
  {B{\oe}uf}}, \ and\ \bibinfo {author} {\bibfnamefont {H.~v.}\ \bibnamefont
  {L\"ohneysen}},\ }\href {\doibase 10.1103/PhysRevB.65.172404} {\bibfield
  {journal} {\bibinfo  {journal} {Phys. Rev. B}\ }\textbf {\bibinfo {volume}
  {65}},\ \bibinfo {pages} {172404} (\bibinfo {year} {2002})}\BibitemShut
  {NoStop}%
\bibitem [{\citenamefont {Al-Kanani}\ and\ \citenamefont
  {Booth}(1995)}]{al-kanani_magnetic_1995}%
  \BibitemOpen
  \bibfield  {author} {\bibinfo {author} {\bibfnamefont {H.~J.}\ \bibnamefont
  {Al-Kanani}}\ and\ \bibinfo {author} {\bibfnamefont {J.~G.}\ \bibnamefont
  {Booth}},\ }\href {\doibase 10.1016/0304-8853(94)01157-5} {\bibfield
  {journal} {\bibinfo  {journal} {Journ. Magn. Magn. Mater}\ }\bibinfo {series}
  {International {Conference} on {Magnetism}},\ \textbf {\bibinfo {volume}
  {140-144}},\ \bibinfo {pages} {1539} (\bibinfo {year} {1995})}\BibitemShut
  {NoStop}%
\end{thebibliography}
%

\end{document}